\documentclass[twocolumn,superscriptaddress,prb,floatfix]{revtex4}
\usepackage[]{graphicx}
\usepackage{color}
\usepackage{soul}
\usepackage{bm}
\usepackage{ulem}

\begin{document}

\title{Possibility of Combining Ferroelectricity and Rashba-like spin splitting \\ in Monolayers of 1T-Type Transition-Metal Dichalcogenides MX$_2$}

\author{Emilie Bruyer}\email{emilie.bruyer@spin.cnr.it} \affiliation{Consiglio Nazionale delle Ricerche
(CNR-SPIN), Via Vetoio, L'Aquila, Italy} \affiliation{Institut des Sciences Chimiques de Rennes, UMR CNRS 6226 - 
Universit\'e de Rennes 1, Campus de Beaulieu, 35042 Rennes, France}

\author{Domenico Di Sante}\email{domenico.disante@physik.uni-wuerzburg.de} \affiliation{Consiglio Nazionale delle Ricerche
(CNR-SPIN), Via Vetoio, L'Aquila, Italy} \affiliation{Institut f\"{u}r Theoretische Physik 
und Astrophysik, Universit\"{a}t W\"{u}rzburg, Am Hubland 97074 W\"{u}rzburg, Germany}

\author{Paolo Barone}
\affiliation{Consiglio Nazionale delle Ricerche
(CNR-SPIN), Via Vetoio, L'Aquila, Italy}
\affiliation{Graphene Labs, Istituto Italiano di Tecnologia, via Morego 30, 16163 Genova, Italy}

\author{Alessandro Stroppa} \affiliation{Consiglio Nazionale delle Ricerche
(CNR-SPIN), Via Vetoio, L'Aquila, Italy}\email{alessandro.stroppa@aquila.infn.it}

\author{Myung-Hwan Whangbo} \affiliation{Department of Chemistry, North Carolina
State University, Raleigh, North Carolina 27695-8204, U.S.A.}

\author{Silvia Picozzi} \affiliation{Consiglio Nazionale delle Ricerche
(CNR-SPIN), Via Vetoio, L'Aquila, Italy}

\date{\today}

\begin{abstract}

First-principles calculations were carried out to explore the possible coupling between spin-polarized electronic states and ferroelectric polarization in monolayers of transition-metal dichalcogenides MX$_2$ (M = Mo, W; X = S, Se, Te) with distorted octahedrally coordinated 1T structures. For d$^2$ metal ions, two competing metal clustering effects can take place, where metal ions are arranged in trimers or zigzag chains. Among these, the former structural distortion comes along with an improper ferroelectric phase which persists in the monolayer limit. Switchable Rashba-like spin-polarization features are predicted in the trimerized polytype, which can be permanently tuned by acting on its ferroelectric properties. The polar trimerized structure is found to be stable for 1T-MoS$_2$ only, while the nonpolar polytype with zigzag metal clustering is predicted to stabilize for other transition-metal dichalcogenides with d$^2$ metal ions.


\end{abstract}


\maketitle

In recent years, phenomena emerging from relativistic electrons in solids have been object of an ever-increasing attention.
Among other effects, relativistic spin-orbit coupling (SOC) provides a mechanism for spin-momentum locking that is appealing for spintronic applications aimed at an all-electric control of spin transport. SOC may lead
to topologically non-trivial insulating states, characterized by the
presence of fully spin-polarized symmetry-protected metallic surface states\cite{SOC}, or
it can give rise to Dresselhaus and Rashba effects in noncentrosymmetric and polar materials\cite{Dresselhaus, Rashba,zunger_nat}. 
More recently, it has been understood that SOC also mediates a spin-valley coupling which is responsible, in noncentrosymmetric materials, for the appearance of valley-contrasting effects, such as valley-selective optical excitations and valley Hall response\cite{valley_rev}.
The lack of inversion symmetry in semi-metallic materials with sizeable spin-orbit coupling may also lead to band structures with non-trivial topology, realizing the so-called Weyl semi-metal phase characterized by pairs of monopoles of Berry curvature in momentum space and topologically protected Fermi arcs in their surfaces\cite{weyl_viewpoint}. 

\begin{figure}[!b]
\centering
\includegraphics[width=\columnwidth,angle=0,clip=true]{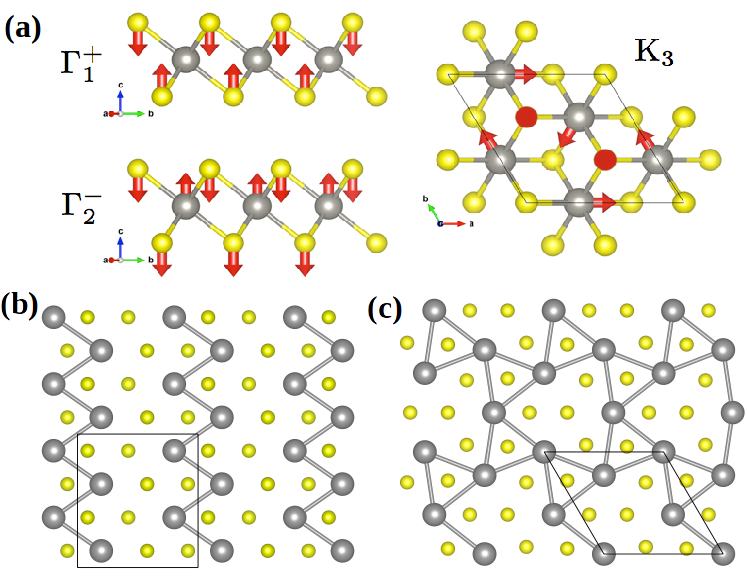}
\caption{(Color online) (a) Distortion modes $\Gamma_1^+$,
$\Gamma_2^-$ and K$_3$ linking the centrosymmetric i-MX$_2$ phase
to the ferroelectric t-MX$_2$ phase \cite{mos2_ferro}. 
The two main clustering effects arising in 1T-MX$_2$ compounds are shown in bottom panels : (b) zigzag chains within the c-MX$_2$ phase 
and (c) trimerized t-MX$_{2}$ phase. For sake of clarity, M-X bonds are not represented here.}
\label{Fig1}
\end{figure}

The class of binary transition-metal dichalcogenides MX$_2$ (where M = Mo, W and X = S, Se, Te) provides an interesting playground where relativistic electronic phenomena can emerge from the relatively strong SOC of $4d$ and $5d$ metal ions. Several polytypes of these layered compounds can be synthetized, and monolayers can be extracted from the bulk using mechanical or chemical exfoliation methods\cite{2d_exp_rev}. Monolayers of transition-metal dichalcogenide MoS$_2$ have been put forward as prototypical valley-active systems, and they have been object of an intense research activity with the aim of establishing their potential appeal for valleytronic applications\cite{valley_rev,valley1,valley2,valley3}. The most common bulk phase of MoS$_2$, the so-called 2H polytype\cite{Wilson1969}, comprises two layers of edge-sharing MoS$_6$ trigonal prisms in its centrosymmetric unit cell (space group $P6_3/mmc$), each layer individually lacking inversion symmetry. MoS$_2$ monolayers obtained via mechanical exfoliation method typically mantain the trigonal prismatic coordination and the acentric (albeit nonpolar) crystal structure; on the other hand, the chemical exfoliation method may result in different octahedral coordinated phases, that can be obtained by distorting the so-called 1T (TiS$_2$-like) polytype, alongside the 2H phase\cite{1TMoS2_jcsc, 1TMoS2_jacs, 1TMoS2_prb, 1TMoS2_tetra, 1TMoS2_tri, 1TMoS2_zigzag, m45, m46, m47, m48, m49}. The ideal 1T-MX$_2$ polytype (henceforth labeled as i-MX$_2$) comprises layers of undistorted edge-sharing
MX$_6$ octahedra and displays inversion symmetry; however, the metal ions in the layers of 1T-MX$_2$ dichalcogenides can undergo a variety of metal-atom clustering, depending on its $d$-electron counting, $d^n$, due to metal-metal bond formation \cite{whangbo-JACS, whangbo-IC}. Among all the distorted 1T-MX$_2$ phases, the case of $d^2$ Mo and W ions is interesting, because the metal ions of 1T-MX$_2$ can form either zigzag-chains (henceforth labeled as c-MX$_2$) or form metal trimers (referred to as t-MX$_2$, see Figure\ref{Fig1}).

Even though the different 1T polytypes are relatively close from a structural point of view, their structural differences affect significantly the properties of relativistic  electrons. Centrosymmetric MX$_2$ monolayers with c-MX$_2$ structure have been  predicted to host 2D topological ($i.e.$ quantum spin-Hall) insulating  phases\cite{Qian2014}. Furthermore, a non-saturating giant magnetoresistance has been reported for non-magnetic semimetal WTe$_2$\cite{cava_nature}, which usually crystallize in a distorted 1T-type acentric crystal structure (orthorhombic space group $Pnm2_1$) showing tungsten zigzag chains. Interestingly, it has been recently suggested that acentric c-WTe$_2$ could be a prototype for a new class of Weyl semimetal, where Weyl points appear as touching points between electron and hole pockets\cite{soluyanov_Nature}.

On the other hand, the distorted t-MoS$_2$ with metal trimerization has been recently predicted to undergo an improper ferroelectric transition with a non-negligible electric polarization\cite{mos2_ferro}.
Ferroelectrics are an interesting subclass of bulk materials lacking inversion symmetry. Due to the presence of a polar axis, ferroelectric semiconductors comprising heavy ions could in principle show SOC-induced Rashba effects (i.e., k-dependent spin splitting in the band structure) which could be tuned by acting on their ferroelectric properties, thus allowing for the integration of a memory functionality brought in by ferroelectricity\cite{SilviaFERSC}. The possibility of a non-volatile fully electrical control of momentum-dependent spin polarization has been recently proposed in ferroelectric semiconductors GeTe and SnTe, in hybrid organic-inorganic halides with perovskite structure and in transition-metal oxide heterostructures\cite{GeTe,SnTe,PNAS,FASNI,ZHONG,BIO_bilayer}. In order to identify candidate materials appealing for realistic applications, however, it is crucial to combine a robust ferroelectricity and semiconducting behaviour with (relatively) high carrier mobility and (relatively) small band-gap. Furthermore, it would be desirable to mantain the ferroelectric properties in the limit of ultrathin films, where typically strong depolarization fields are experienced by conventional oxide ferroelectrics leading to a strong suppression of ferroelectric polarization\cite{ghosez_ultrathinFE}.
Due to its improper origin, the ferroelectric phase in trimerized MoS$_2$ has been predicted to be unexpectedly robust up to the monolayer limit, at the same time mantaining a small band-gap of the order of 1 eV\cite{mos2_ferro}. It represents, therefore, an ideal candidate where the Rashba-like spin-splitting effects and their interplay with ferroelectric polarization could be explored.


In this work we explore the possible coexistence of ferroelectricity and spin-polarized
states in distorted 1T-MX$_2$ phases for M = Mo, W and X = S, Se, Te. Some of these phases are already met at
ambient conditions, and are also highly tunable by a moderate degree of strain
\cite{Qian2014}. 
In addition, two-dimensional (2D) heterostructures composed of
different polymorphs of MX$_2$ might introduce more functionality in new devices \cite{Radi2011}. 
While all these properties open new routes for multifunctional
materials, a deep understanding and characterization of the full class of MX$_2$
is strongly needed. In this context we explore two important questions
concerning the t-MX$_2$ layers on the basis of density functional calculations:
one is to quantify their ferroelectric (FE) and SOC related properties among
several t-MX$_2$ layers, and the other is to examine whether the t-MX$_2$
structure is sustainable in view of the fact that the c-MX$_2$ structure, which
has not been predicted to display a FE polarization, is an alternative polymorph of t-MX$_2$.

We carry out density functional calculations using the projector augmented wave
(PAW) method implemented in VASP using the PBE functional \cite{paw,vasp,pbe}.
Atomic positions and in-plane lattice parameters of MX$_2$ monolayers 
were optimized until the residual forces were smaller than 0.001 eV/\AA\, with the
plane-wave cutoff energy of 500 eV and a set of 10$\times$10$\times$1 k-points
for the irreducible BZ. Repeated image interactions were made negligible by
including a vacuum layer of 20 \AA\, in the simulations. The FE polarization is
calculated using the Berry's phase method \cite{berry}, and the volume factor in
the estimation of monolayer polarization is $V=Sc/2$, where $S$ refers to the in-plane area of the monolayer unit cell and
$c$ to the bulk lattice constant of the 1T-MX$_2$ phase, as previously used in ref. \cite{mos2_ferro}. In order to properly treat relativistic effects, SOC is self consistently taken into account and dipole corrections are included.


We start our discussion by reviewing the structural transition from the i-MX$_2$ to the polar t-MX$_2$ structure. The main distortion mode displays a $K_3$ symmetry, corresponding to a trimerization of metal ions resulting in a $\sqrt{3}\times\sqrt{3}$ superstructure. Its origin can be roughly understood in terms of a Peierls instability of hidden one-dimensional bands. In the ideal 1T structure, in fact, the band structure can be viewed as a superposition of three 1D bands arising from the strong direct hybridization between $t_{2g}$ states along three chains containing the equatorial planes of edge-sharing octahedra (i.e., $xy, yz, zx$ in the local reference frame of the MX$_6$ octahedra). Due to the directional dependence of $t_{2g}$ orbitals, metal-metal intra-orbital interactions in each edge-sharing octahedral chains are strong (being $\sigma$ in nature), while inter-chains and inter-orbital intra-chains interactions are weak. In the $d^2$ case, each degenerate hidden band is 1/3-filled, resulting in a metallic state that is susceptible to structural distortions in the presence of sizeable electron-phonon interaction which may lower the crystal symmetry and open a band gap at the Fermi level\cite{whangbo-JACS, whangbo-IC}. {\color{black}In the trimerized structure, each chain displays a sequence of long-short-short metal-metal bond lengths, leading to a complete removal of degeneracies in each hidden 1D band and to the opening of a gap of the order of 1 eV. }

In this picture, only the $t_{2g}$ orbital states of the metal ions M play a fundamental role in the mechanism of the structural instability\cite{mos2_ferro}. Nonetheless, two additional distortions can be identified at the structural transition, namely a $\Gamma_1^+$ and a $\Gamma_2^-$ mode, as shown in Fig. \ref{Fig1}. While the former corresponds to a deformation of the MX$_6$ octahedra, inducing a trigonal crystal-field effect in the band-structure, the $\Gamma_2^-$ mode is responsible for the inversion-symmetry breaking leading to the ferroelectric state. As detailed in Ref.\cite{mos2_ferro}, such polar mode can be viewed as a secondary order parameter, nonlinearly coupled to the primary order parameter describing the condensation of the $K_3$ mode and, as such, describing an improper ferroelectric transition. This improper ferroelectricity is expected to develop as soon as the primary ordering occurs, being weakly affected by the presence of depolarization fields which would only slightly harden the polar mode\cite{mos2_ferro, improper_fennie}. 

Both the secondary $\Gamma_1^+$ and $\Gamma_2^-$ modes involve primarily a change in the M-X bonds, which are highly covalent due to the hybridization between the chalcogene $p$ and the metal $e_g$ states. Consistently, their amplitudes, tabulated in TABLE \ref{tab1}, increase by increasing the covalency of the M-X bond, {\it i.e.} on
going from S to Se to Te, while they are hardly affected by a change in the
transition metal site. The high covalency of M-X bonds 
can be inferred from the highly anomalous Born-effective charges\cite{mos2_ferro} and by comparing the ferroelectric polarization calculated by using the point charge model
(assigning to each atom its nominal valency charge) with that obtained in the framework of the Berry Phase formalism.
While the former, suitable for ionic systems, gives unreasonably large values for the
polarization, {\it i.e.} $\sim 4  \mu C/cm^2$, the latter leads to much smaller
values, of the same order of magnitude for all compounds, {\it i.e.}
$\sim 0.1-0.2  \mu C/cm^2$\cite{footnote}.
Such small dielectric polarization, implying smaller depolarization fields, together with the improper origin of the ferroelectric state, is highly
desirable for 2D ferroelectrics applications,  as the 
ferroelectric polarization is found to persist even in the ultrathin (monolayer) film limit\cite{ultrathin}.

\begin{table*}
\centering
\caption{In-plane lattice parameters (\AA), average M-X and M-M bond lengths in t-MX$_2$ structure (\AA), amplitudes of $\Gamma_1^+$,
$\Gamma_2^-$ and K$_3$ distortion modes (\AA), energy difference between i-MX$_2$ and t-MX$_2$ structures per formula unit (meV/fu), 
total ferroelectric polarization calculated using the Point Charge Model and the Berry's Phase formalism ($\mu C/cm^2$), energy gap (eV) and estimated energy difference between t-MX$_2$ and c-MX$_2$ structures (meV/fu). The two M-M distances listed here refer to the shortest M-M distances before and after the metal-atom clustering. These numbers are written without and with parenthesis, respectively.} 
\label{tab1}
\begin{tabular}{p{1cm}ccccccccccccc}
\hline\hline
     & a     & d(M-X) & d(M-M) & & & $\Gamma_1^+$   & $\Gamma_2^-$  & K$_3$  & $\Delta E$   & P$_{PCM}$      & P              & E$_g$  & $\Delta E$$_{t-c}$ \\
     & (\AA) & (\AA)  & (\AA)  & & & (\AA)          & (\AA)         & (\AA)  & (meV/fu)     & ($\mu C/cm^2$) & ($\mu C/cm^2$) & (eV)   & (meV/fu) \\
\hline
MoS$_2$ & 5.47 & 2.43 & 3.75 (3.02) & & & 0.089 &  0.055     & 0.631  & 242 & 4.50 & 0.23 & 0.74 & -22.0 \\
MoSe$_2$ & 5.83 & 2.55 & 3.37 (3.12) & & & 0.090 &  0.067     & 0.745  & 327 & 4.81 & 0.15 & 0.68 & +57.8 \\
MoTe$_2$ & 6.17 & 2.74 & 3.53 (3.28) & & & 0.104 &  0.081     & 0.858  & 312 & 4.75 & 0.10 & 0.58 & +146.2 \\
WS$_2$ & 5.64 & 2.43 & 3.26 (3.04) & & & 0.043 &  0.051     & 0.641  & 315 & 4.40 & 0.18 & 0.77 & +36.5 \\
WSe$_2$  & 5.85 & 2.56 & 3.38 (3.13) & & & 0.071 &  0.064     & 0.739  & 369 & 4.56 & 0.21 & 0.66 & +146.9 \\
WTe$_2$ & 6.22 & 2.75 & 3.59 (3.31) & & & 0.083 &  0.081     & 0.858  & 327 & 4.73 & 0.11 & 0.55 & +280.0 \\
\hline
\hline
\end{tabular}
\end{table*}

\begin{figure}[h!t]
\centering
\includegraphics[width=\columnwidth,angle=0,clip=true]{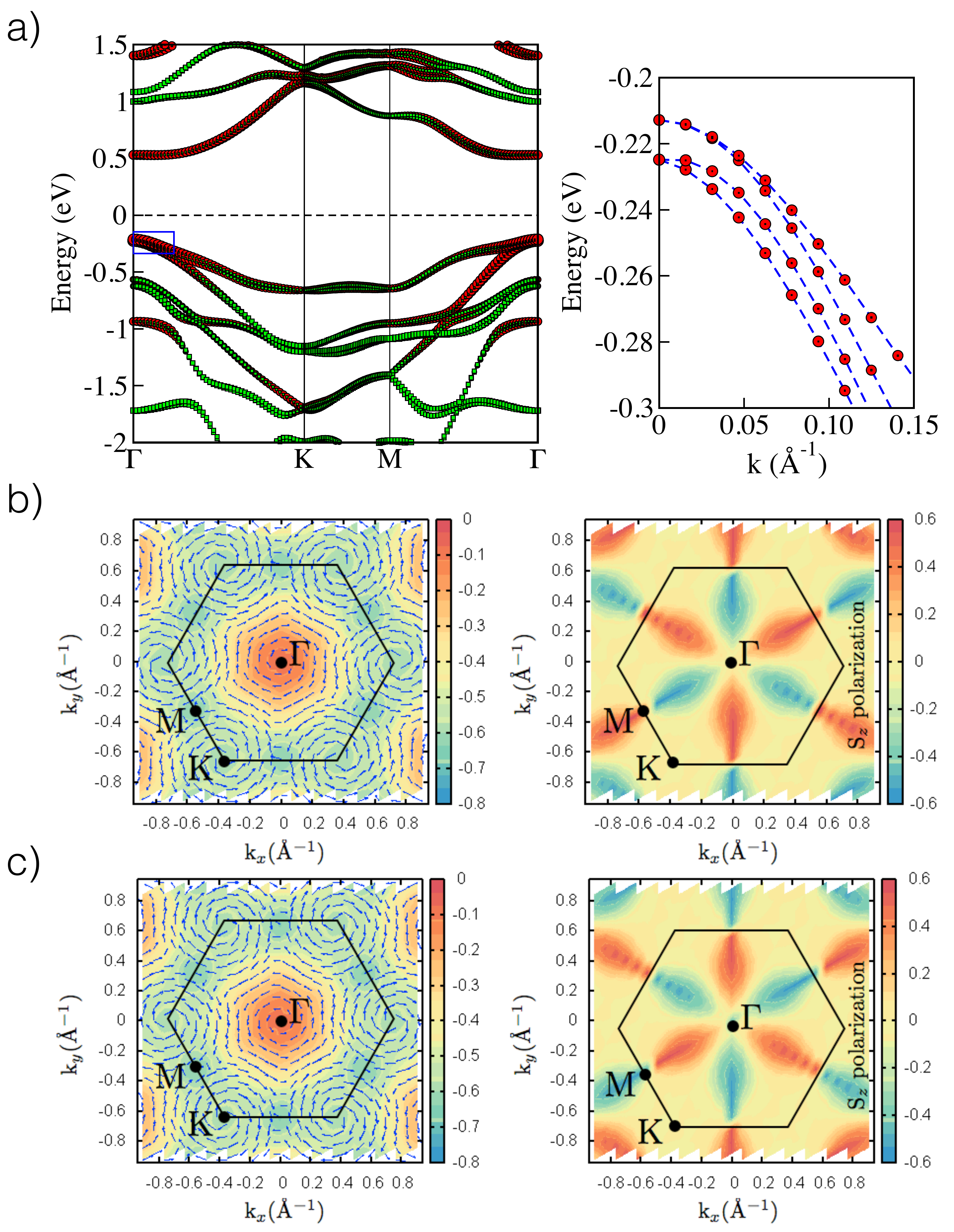}
\caption{(Color online) a) Band dispersions for t-MoS$_2$ including spin-orbit coupling.
Red circles highlight characters from transition metal d orbitals, while green circles
refer to chalcogenides p orbitals. Zoom on the energy bands at the $\Gamma$ point (up-right panel) evidences the SOC-induced band dispersion.
b) In-plane (left panels) and out-of-plane (right panels) spin-textures for
t-MoS$_2$'s valence band. b) and c) panels refer to opposite directions of the
ferroelectric polarization. For left panels, the color scale refers to valence
band's energy (eV) with respect to the Fermi level, highlighting the $C_{3v}$
symmetry, and in-plane spin directions are represented by blue arrows, while for
right panels the color scale quantifies to the out-of-plane S$_z$ spin
polarization.}
\label{Fig2}
\end{figure}

We discuss now the effects of SOC in the electronic structure of the ferroelectric t-MoS$_2$. The presence of a relatively heavy element like molybdenum justifies our aims of fulfilling this aspect neglected in Ref. \cite{mos2_ferro}. 
In Fig.\ref{Fig2}a we show the band
structure for t-MoS$_2$ with SOC included. When looking at the orbital character distribution, we
recognize that the states around the gap {\color{black} at $\Gamma$} present a strong transition metal Mo
weight (red color), while significant S contributions occur at higher energies
with respect to the Fermi level {\color{black} and/or away from the $\Gamma$ point}. This reflects the fact that, with d$^2$
electron count for the Mo$^{4+}$ ion, only the low-lying portion of the
t$_{2g}$ bands is occupied (consistently with the Peierls-like scenario), while the whole set of empty e$_{g}$ bands, where
the S contributions are strong, lie well above the t$_{2g}$-group bands. 
Inclusion of SOC removes the spin degeneracy, inducing nontrivial spin-textures, even though the observed spin-splitting are rather small. In Fig.\ref{Fig2}b we report the in-plane and out-of-plane spin-textures for the
valence bands of the t-MoS$_2$ monolayer along the whole 2D Brillouin Zone.
Around the $\Gamma$ and K points, spins are circularly rotating with in-plane spin components perpendicular to the wave
vectors as in a Rashba
system (left panel), consistently with the polar symmetry of the ferroelectric t-MoS$_2$\cite{zunger_nat}. The out-of-plane spin component, on the other hand, is nonvanishing
only along the $\Gamma$M lines (right panel). The weakness of the SOC-induced spin-splitting effects can be ascribed to the substantial robustness of the hidden 1D bands arising from $t_{2g}$ states and forming the band structure around the Fermi level. Generally speaking, the Rashba interaction, which couples spin to an axial orbital vector which is odd under inversion symmetry, arises from hopping processes that are odd under inversion in the monolayer plane\cite{winkler_book, macdonald_prb2013}. The ferroelectric phase of t-MoS$_2$ is however expected to open new covalency channels mostly between the metal $e_g$ and chalcogene $p$ states, being less effective in the $t_{2g}$ manifold where the dominant interaction is the direct $\sigma$-like hybridization between $d_{xy}$, $d_{yz}$ and $d_{zx}$ orbitals. 
In this respect, we notice that, due to the nontrivial band structure of the distorted t-MX$_2$ structure, the standard Rashba Hamiltonian\cite{Rashba, winkler_book} cannot be used either to model the observed spin splittings and spin textures or to assess the strength of the Rashba effect.
Nevertheless, a full electric control of the spin-texture of the valence band maximum is predicted when acting on the
ferroelectric polarization. As shown in Fig.\ref{Fig2}c, a polarization reversal leads to a full switching of the spin-textures, both for in-plane and out-of-plane spin components.

\begin{figure*}
\centering
\includegraphics[width=\textwidth,angle=0,clip=true]{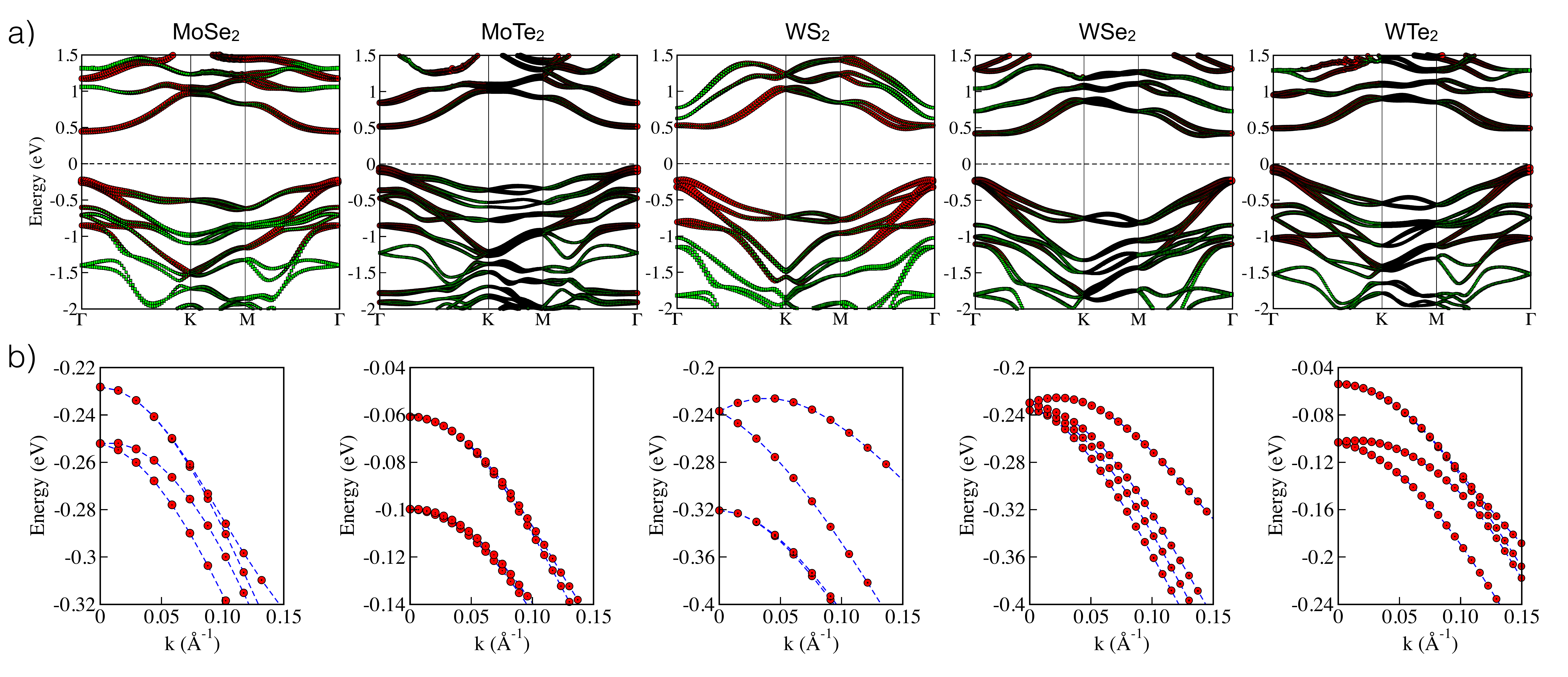}
\caption{(Color online) Relativistic bandstructures
for t-MX$_2$ compounds (MoSe$_2$, MoTe$_2$, WS$_2$, WSe$_2$ and WTe$_2$ from left to right) along the
irreducible 2D BZ lines. Red circles highlight characters from transition metal
d orbitals, while green circles refer to chalcogenides p orbitals. {\color{black} Lower panels show a zoom on the energy bands around the $\Gamma$ point.} }
\label{Fig3}
\end{figure*}

A direct comparison between band dispersions with (Fig.\ref{Fig2}a) and without
(Fig.2b of Ref.\cite{mos2_ferro}) SOC for the t-MoS$_2$ layer shows that the
overall results of SOC-induced effects do not lead to substantial differences.
To see the effect of SOC enhancement, we extended our study to the t-MX$_2$ (M =
W; X = Se, Te) layers. The band structures calculated for these
layers along the 2D Brillouin Zone are plotted in Fig.\ref{Fig3}, and the band
gaps are summarized in TABLE \ref{tab1}.
{\color{black} Despite the SOC-induced spin-splittings are not much enhanced, the combined effect of ferroelectricity and SOC persists for
all t-MX$_2$ layers, and it becomes stronger as the ligand X changes from Te
to Se to S and as the metal changes from Mo to W, as shown in Fig. \ref{Fig3}b).} 
As for the orbital character of the band structure, we
note from Fig. \ref{Fig3} that the low energy features around the Fermi level {\color{black}  at $\Gamma$} in
all cases show a predominant weight from the transition-metal {\it d} orbitals
(as for t-MoS$_2$), both in valence and conduction bands. Increasing the anionic size induces a
decrease in the gap, because it leads to longer metal-metal bonds, implying a decrease of the direct hopping interaction between $t_{2g}$ states (which scales roughly as $b^{-5}$, where $b$ is the M-M bond length), at the same time {\color{black} enhancing the role of hybridization between the M-$d$ and X-$p$ states, as can be clearly seen from the evolution of band characters in Fig.\ref{Fig3}}. The
size of the gap is an important issue for the ferroelectric switchability. In
our case, we found the gaps to be about 0.5-0.7 eV, similar to standard
ferroelectric semiconductors, and greater than those predicted in the recently studied c-MX$_2$ layers
\cite{Qian2014}. Additionally, a non-inverted band gap is detected for all
t-MX$_2$ layers, suggesting that the band-structure topology remains trivial in this case, as opposed to monolayers with
c-MX$_2$ structure\cite{Qian2014}. {\color{black} Indeed, the explicit calculation of $Z_2$ topological invariants using the hybrid Wannier charge centers\cite{Soluyanov_PRB2011}  confirms the trivial character of the t-MX$_2$ band-structure. } 

\begin{figure}[!t]
\centering
\includegraphics[width=\columnwidth,angle=0,clip=true]{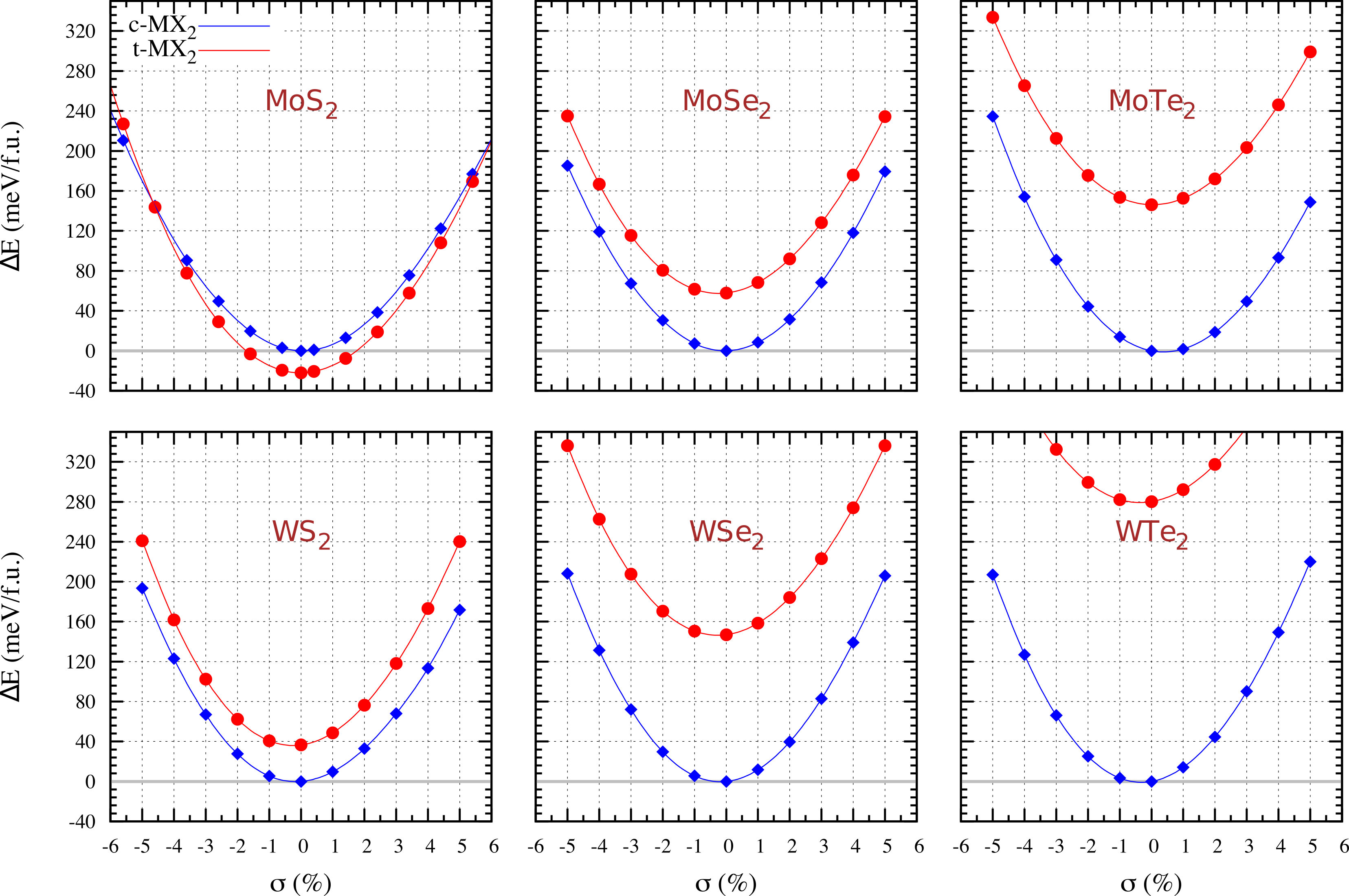}
\caption{(Color online) Relative stabilities (meV/fu) between t-MX$_2$ (red dots) and
c-MX$_2$ (blue dots) structures as a function of applied in-plane strain for all
1T-MX$_2$ compounds.}
\label{Fig4}
\end{figure}

Since the 1T-MX$_2$ layers with
d$^2$ metal ions can have the t-MX$_2$ or the c-MX$_2$ structure and since only
the t-MX$_2$ structure can be ferroelectric, it is necessary to examine how
their relative stabilities depend on the M and X atoms. 
In general, the pattern of metal-atom
clustering in the 1T-MX$_2$ layer depends on the balance between two main competing factors, namely the overall energy gain arising from the direct M-M bonding interaction and the energy loss arising from the lattice strain induced by the metal-atom clustering. Experimentally, the
t-MX$_2$ structure has been observed in LiVO$_2$ and 1T-MoS$_2$ \cite{LVO_mrb1992, 1TMoS2_jcsc, 1TMoS2_tri}, for which the M-X bonds
(i.e., V-O, Mo-S) are short, while the c-MX$_2$ structure is found for
1T-MoTe$_2$, 1T-WTe$_2$, and M$_{0.5}$Nb$_2$Se$_4$, for which the M-X bonds (i.e.,
Mo-Te, W-Te, Nb-Se) are long \cite{MoTe2_zz,MNS_zz}. 
Roughly speaking, shorter M-X distances imply less polarizable M-X bonds, and hence an increase of the lattice strain;  on the other hand, it has been argued that the energy gain of the metal-ion zigzag clustering is reduced when compared to metal trimerization as the width of the $t_{2g}$-block bands increases, i.e., when the M-M distance is reduced\cite{whangbo-IC}. 
Since the metal trimer (zigzag chain) formation requires small (large) displacements of M from the ideal structure i-MX$_2$, thus minimizing (maximizing) the lattice strain,
it has been suggested that the
t-MX$_2$ structure is favorable for 1T-MX$_2$ systems with small M-X bonds, but
the c-MX$_2$ structure for those with long M-X bonds \cite{whangbo-IC}. 

To verify this suggestion, we carry out the structure optimization for the
t-MX$_2$ and c-MX$_2$ (M = Mo, W  X = S, Se, Te) layers as a function of their 2D
unit cell parameter $a$. This was achieved by building a large supercell which can accomodate both types of distortions;  an in-plane strain was mimicked through the modification of the in-plane lattice parameter $a$ while the $a/b$ ratio was kept constant.The relative energies of the two structures thus
determined are summarized in Fig. \ref{Fig4} {\color{black} and in Tab. \ref{tab1}}. The t-MX$_2$ structure is more
stable than the c-MX$_2$ only for 1T-MoS$_2$, which has the shortest M-X bond.
For all other 1T-MX$_2$ with the M-X longer than the Mo-S bond, the c-MX$_2$
structure is more stable than the 1T-MX$_2$ structure, with an energy gain increasing as the size of the X ion (and hence of the M-X bond length) increases. That being said, the energy barrier for WS$_{2}$ and MoSe$_{2}$ is not very large, suggesting that these metastable phases may be obtained under appropriate growth conditions. Our results therefore show that out of
all 1T-MX$_2$ phases, one can expect the combined effect of ferroelectricity and
SOC-induced Rashba spin splitting in 1T-MoS$_2$; despite being the effect of SOC not very strong in this compound, one can expect to obtain full control and switchability of its spin texture under an applied electric field.


Experimentally, separated MX$_{2}$ layers of 1T- and 2H-MX$_{2}$ phases can be obtained by chemically exfoliating them \cite{m45, m46, m47, m48, m49}. In the typical intercalation-exfoliation method, the binding between adjacent MX$_{2}$ layers is loosened by first intercalating alkali metals ($e.g.$, Li, Na, K) between them, and then the layers are separated by solvents\cite{m45,m46,m47,m48}. The intercalated metals exist as their cations ($i.e.$, Li$^{+}$, Na$^{+}$, K$^{+}$), so that the electrons lost by these metals are accommodated by the d-block bands of the layers, thereby changing the $d$-electron count on the metal $M$. This situation will remain the same even for the exfoliated MX$_{2}$ layers, because their surfaces will be covered by alkali cations. The metal-atom cluster patterns of MX$_{2}$ layers depend on the $d$-electron count on the metal \cite{whangbo-JACS,whangbo-IC}. Then, for the separated MoS$_{2}$ layers from 1T-MoS$_{2}$ by the interaction-exfoliation procedure, the $d$-electron count on Mo would be different from $d$$^{2}$ found in the bulk 1T-MoS$_{2}$ so that their metal-atom clustering pattern will differ from that found in the bulk 1T-MoS$_{2}$. However, the isolated MX$_{2}$ layers retaining the properties of those in the bulk can be re-generated by annealing them at elevated temperature \cite{m48}, probably because the alkali cations on their surface are removed in the form of alkali metal atoms. It is also noted that separated MX$_{2}$ layers retaining the properties of the bulk MX$_{2}$ are obtained by using chlorosulfonic acid instead of alkali metals\cite{m49}. Therefore, it should be possible to prepare separated t-MoS$_{2}$ layers and verify their ferroelectric properties predicted in the present work.


In conclusion, we found that all t-MX$_2$ layers with d$^2$ metal ions show a spontaneous dielectric
polarization in the range 0.1--0.2 $\mu C/cm^2$ with band gaps around $0.6$ eV, a
reasonably large value comparable with ferroelectric semiconductors with non
trivial spin-textures \cite{SilviaFERSC}. Due to its improper origin, ferroelectric polarization persists in the ultrathin film limit of t-MX$_2$ monolayers. Spin splitting and Rashba-like spin-polarization features are predicted, which appear to be reversed when switching the ferroelectric polarization, allowing full control of the spin degrees of freedom via an external electric field. However, the t-MX$_2$ structure is
energetically stable only for 1T-MoS$_2$ while for all other
1T-MX$_2$ phases, the nonpolar c-MX$_2$ structure is more stable than the polar
t-MX$_2$ structure. We therefore hope that our work will stimulate more experiments in ferroelectric 1T-MoS$_2$ in order to explore relativistic phenomena and spin-splitting effects in its band structure.

We acknowledge CINECA for providing us computational resources (IsC33\_TRAMDIS Project) and Henry
cluster from North Carolina State University. D. D. S. and A. S.
acknowledge the CARIPLO Foundation through the MAGISTER Project No. Rif.
2013-0726. D. D. S. acknowledges financial support from the German
Research Foundation (DFG-SFB Project No. 1170).


\begin{thebibliography}{50}
\bibitem[{foo()}]{footnote}
\bibinfo{note}{Our result for t-MoS$_2$ differs slightly from that recently
  reported in ref.\cite{mos2_ferro} (P $\sim 0.18 \mu C/cm^2$); this
  discrepancy is simply due to the fact that our values were normalized with a
  different volume factor from theirs.}
\end{thebibliography}
\end{document}